# Directed Self-Assembly of Quantum Dots in a Nematic Liquid Crystal

Rajratan Basu and Germano S. Iannacchione*

*Order-Disorder Phenomena Laboratory, Department of Physics,*

*Worcester Polytechnic Institute, Worcester, MA 01609, USA*

December 6, 2008

Self organizing anisotropic nematic liquid crystals (LCs) induce self-assembly on quantum dots (QDs) to form one-dimensional chains along the nematic director. Spatial ordering of QDs, achieved in the nematic LC matrix, can be controlled in a preferred direction on application of electric fields, gaining better self-assembly. Once the field goes off, the LC+QD system relaxes back to the original state, revealing the intrinsic dynamics. Due to the dielectric anisotropy of the system, this dynamic response can be captured by studying dielectric relaxation. The studied dynamic response reveals insights on the field-induced self-assembly mechanism and the stability of the LC+QD system.

*Electronic address: gsiannac@wpi.edu

Controlled self-assembly of semiconductor Quantum Dots (QDs) holds great promise for numerous applications, such as next generation photonic devices, QD displays, biomedical imaging [1,2,3,4,5,6,7,8,9] and, perhaps, solid-state quantum computation. Electrochemical self-assembly of QDs on a chosen substrate is one of the most efficient techniques to form highly ordered QD-aggregates. However, this technique, like others, does not allow manipulating the QD-aggregates in a preferred direction after the completion of the self-assembly process. It has been demonstrated that nano-materials like nanotubes or nano-rods can be organized in a nematic *liquid crystal* (LC) environment [10,11,12]. In all such cases, self organizing anisotropic LC molecules induce alignment on the nano-size guest particles along the *director* (average direction of LC molecules) due to the reduction of excluded volume [13] and, the LC phases (nematic or smectic) act like patterned templates for alignment. Recently, it has been manifested that highly spatial ordering of QDs can be achieved in a smectic LC environment [14]. But, unlike the smectic phase, the nematic phase enables controlled director orientation using well-established methods of LC alignment, such as patterned-electrode-surface, electric field, and magnetic field. Thus, the spatial ordering of QDs achieved in a nematic LC environment can also be manipulated on application of electric fields or magnetic fields. In this letter, we report the dynamic response of the average dielectric constant, $\bar{\varepsilon}$ for an LC+QD system, to probe the electric field induced self-assembly mechanism of cadmium sulfide (CdS) QDs in nematic 4-Cyano-4′-Pentylbiphenyl (5CB) LC media.

The LC system shows dielectric anisotropy due to the structural anisotropy. For a positive dielectric anisotropic system, planar configured molecules, being perpendicular to the measuring field, show smallest dielectric constant, $\varepsilon = \varepsilon_\perp$. When the director rotates, the dielectric constant increases until the system is homeotropically aligned (parallel to the measuring field) and, the dielectric constant reaches its largest value, $\varepsilon_{||}$. However, the values of $\varepsilon_\perp$ and $\varepsilon_{||}$ depend on the frequency of the measuring field. CdS quantum dots are individually spherical in shape, thus, they do not show dielectric anisotropy in bulk. In an LC+QD system, the total excluded volume has to be the smallest in order to minimize the free energy [13]. The only way the system can do this, is to form QD-strings along the LC-director instead of isotropic distribution of QDs in LC. The number of the QD-strings depends on the concentration of QDs in the LC media. Thus, these zero-dimensional QDs aggregate to form one-dimensional arrays when they are embedded in a nematic-LC environment, as shown in Fig. 1. These one-dimensional QD-aggregates being anisotropic in structure show dielectric anisotropy.

A small amount (1 wt %) of CdS quantum dot sample (UV absorption peak: *361 nm* and diameter- *2.3 nm* in toluene solvent) [15] was dispersed in 5CB ($T_{nematic\text{-}isotropic} = 35^{o}C$) and the mixture was ultrasonicated for 5 hours to achieve well dispersion. The mixture, then, was heated to evaporate the toluene and degassed under vacuum. The mixture was filled into a cell ($5 \times 5\ mm^2$ indium tin oxide (ITO) coated area and $20\ \mu m$ spacing) [16] by capillary action. The patterned-electrode-surface inside the LC cell imposes the planar alignment to the nematic director. Empty LC cells were measured first in order to extract the absolute $\bar{\varepsilon}$. The electric field induced dielectric response of the nematic director also depends on cell configuration; for comparisons, the same type of cells was used for both pure 5CB and 5CB+CdS.

After the 5CB+CdS sample was loaded into the cell, an external ac electric field pulse, $E_{ac}$ (1 MHz) of 30 seconds duration was applied across the cell at magnitudes ranging from 0 - 250 kV/m. The reason for applying the ac field (not dc) is to avoid the affect of ion migration on the dielectric relaxation measurements. Once $E_{ac}$ was turned off (at t = 0 sec), isothermal average dielectric ( $\bar{\varepsilon}$ ) measurements were carried in the nematic (T = 25$^{o}$ C) phase as a function of time. See Fig. 2. The dielectric measurements were performed by the ac capacitance bridge technique [[17],[18],[19]], operating with a probing field far below the reorientation threshold field and at 100 kHz frequency.

Field induced director orientation occurs when the torques, due to the external electric field, overcome the elastic interactions between LC molecules and, being embedded in the LC matrix, one-dimensional QD-chains follow the director reorientation. Soon after the field goes off, these restoring forces, between the planar surface state and LC director, drive the system back to the original state. Figure 2 shows the average complex dielectric constant, $\bar{\varepsilon}$ (100 kHz) as a function of time after $E_{ac}$ was tuned off for 5CB+CdS sample. In the nematic phase (T = 25$^{o}$ C), QD-arrays and LC molecules cooperatively relax back to the planar orientation after the field goes off. The field-saturated dielectric constant, $\bar{\varepsilon}_{max}$ ( $\bar{\varepsilon}$ at t = 0, from Fig. 2) for each relaxation is plotted as a function of $E_{ac}$ in Fig. 3a and is directly associated with the director orientation. The value of $\bar{\varepsilon}_{max}$ starts to increase above $E_{ac}$ = 20 kV/m for both pure 5CB and 5CB+CdS samples, confirming the director reorientation from planar to homeotropic. Before the field-induced reorientation for the LC+QD system, one-dimensional QD-arrays, being perpendicular to the measuring field, contribute their average $\varepsilon_{\perp}$ to the average dielectric constant of the system. Figure 3a shows that the average dielectric constant of the composite system increases by an amount $\Delta\varepsilon_1 = 0.08$. After the saturation point, when the system is fully reoriented parallel to the field, QD-arrays also show homeotropic alignment, contributing their average $\varepsilon_{||}$ ($> \varepsilon_{\perp}$) to the system. The difference in $\bar{\varepsilon}_{max}$ between pure 5CB and 5CB+CdS after the saturation is given by $\Delta\varepsilon_2 = 0.15$ (87.5 % increase), shown in Fig. 3a. The significant difference between $\Delta\varepsilon_1$ and $\Delta\varepsilon_2$ ($\Delta\varepsilon_2 > \Delta\varepsilon_1$) confirms that the spherical QDs form anisotropic one-dimensional arrays that follow the nematic director reorientation on application of $E_{ac}$. If the QDs were to stay in the LC matrix without forming the arrays, one would expect $\Delta\varepsilon_1$ to be equal to $\Delta\varepsilon_2$. The intermediate step found in $\bar{\varepsilon}_{max}$ (Fig. 3a) for 5CB+CdS system indicates that a small amount of QDs present in a nematic LC induces local quenched random disorders. Strong enough $E_{ac}$ allows the system to achieve better nematic ordering, compensating the disorder effect. This field-induced better nematic order enhances the self-assembly of QDs, resulting a dramatic increment in $\bar{\varepsilon}_{max}$ with increasing $E_{ac}$ after the intermediate step.

The same experiment was repeated in the isotropic phase, at T = 45 $^{o}$C. Due to the absence of elastic interactions in the isotropic phase, the LC molecules no longer form long range orientation order. So, it is expected to have no field induced director reorientation for pure 5CB in the isotropic phase, as also experimentally observed in Fig. 3b. The value of $\bar{\varepsilon}_{max}$ of the LC+QD system does not change on application of $E_{ac}$, indicating that the QDs do not seem to form self-assembled arrays in the isotropic phase, seen in Fig. 3b. Also the dielectric constant of the CdS solution (2 mg/cc in toluene) under the same experimental condition does not depend on $E_{ac}$, confirming that CdS nanocrystals in bulk are not field responsive to form arrays.

Dielectric relaxation curves for 5CB and 5CB+CdS composite were fitted according to a single exponential decay function $f(t) = \bar{\varepsilon}_1 e^{(-t/\tau)} + \bar{\varepsilon}_0$ with a typical regression coefficient of R = 0.9976. The inset in Fig. 2 shows the linear dependency of $\bar{\varepsilon}$ with logarithmic time scale. Here, $\tau$ is the relaxation decay time, $\bar{\varepsilon}_0$ is the average base dielectric constant, and $\bar{\varepsilon}_1$ is the field-induced average dielectric constant. Thus, the field-saturated average dielectric constant, $\bar{\varepsilon}_{max} = \bar{\varepsilon}_0 + \bar{\varepsilon}_1$. The values for the three fitting parameters, $\tau$, $\bar{\varepsilon}_0$, and $\bar{\varepsilon}_1$ as a function of $E_{ac}$ are shown in Fig. 4. The relaxation time for pure 5CB and 5CB+CdS decreases as $E_{ac}$ increases and saturates at a higher field, which is consistent with the behaviors of $\bar{\varepsilon}_{max}$ shown in Fig. 3a. For $E_{ac}$ larger than the saturation point, the composite system relaxes back slower than pure 5CB. It is possible that the presence of QD-arrays increases the local viscosity and allows the system to relax slower. The fitting parameter, $\bar{\varepsilon}_1$, recovers the shape of $\bar{\varepsilon}_{max}$ for pure 5CB and 5CB+CdS, indicating that the single exponential decay function works well for fitting the dielectric relaxation curves for this case. The difference in $\bar{\varepsilon}_1$ between 5CB and 5CB+CdS after the saturation point is $\Delta\varepsilon = 0.0698$. See Fig. 4. As expected, this value of $\Delta\varepsilon$ has been found to be very close to the value of $\Delta\varepsilon_2 - \Delta\varepsilon_1$ from Fig. 3a.

A rough estimation of number of QD-chains and number of QDs in a chain can be done based on the following simple model. For the QD-chains, as explained earlier, the homogeneous configuration contributes $\varepsilon_{||}$ and the homeotropic configuration contributes $\varepsilon_{\perp}$ to the average dielectric constant of the host system. Thus, $\Delta\varepsilon = \varepsilon_{||} - \varepsilon_{\perp}$. Quantum dots were initially dispersed isotropically in toluene solvent. One can extract the average contribution, $\delta\varepsilon_0$, of QDs to the dielectric constant of any isotropic media once the average dielectric constants for toluene+QD and pure toluene are known. Now, if $n$ number of QDs form a chain, then one can write $\Delta\varepsilon \cong n\ \delta\varepsilon_0 - \delta\varepsilon_0$ *or*, $n = (\Delta\varepsilon/\ \delta\varepsilon_0) - 1 \approx 2000$. Also, the total number of QDs in the capacitive cell is known, $N \approx 70 \times 10^{12}$. Thus, the total number of chains can be estimated, $m = n/N = 35 \times 10^{9}$.

In summary, the dynamics of 5CB+CdS system has been probed by studying average dielectric response to understand the stability of this system. The results clearly demonstrate that the nematic phase imposes self-assembly on QDs to form one-dimensional arrays. A strong electric field improves the self-assembly in a preferred direction. To strengthen this interpretation, the isotropic phase of the system does not respond to $E_{ac}$, further indicating that only the nematic phase of LC induces this self-assembly on CdS nanocrystals. Future work involves optoelectric studies of field-induced fluorescence spectra for different sizes of QDs in nematic LC media.

**Figure captions:**

**FIGURE 1:** (Color online) **a)** Schematic diagram of electrode-surface-induced homogeneous alignment of nematic LC molecules (ellipsoidal), and QD self-assembly (spherical) in the nematic matrix. **b)** Schematic diagram of electric field induced homeotropic alignment of nematic LC molecules, and homeotropically directed one-dimensional QD arrays.

**FIGURE 2:** (Color online) Dynamic response of the average dielectric constant $\bar{\varepsilon}$ for the 5CB+CdS system in the nematic phase (T = 25°C) after $E_{ac} = 0$; the inset (same main graph axes) represents the same relaxation in log-time scale to show the single exponential decay. The legend represents the magnitude of $E_{ac}$ (1 MHz) in kV/m.

**FIGURE 3:** (Color online) **a)** Field-saturated dielectric constant, $\bar{\varepsilon}_{max}$ ( $\bar{\varepsilon}$ at t = 0) as a function $E_{ac}$ for pure 5CB and 5CB+CdS in the nematic phase (T = 25°C). The downward arrow indicates the intermediate step for 5CB+CdS. Lines represent guide to the eye; **b)** Field-saturated dielectric constant, $\bar{\varepsilon}_{max}$ ( $\bar{\varepsilon}$ at t = 0) as a function $E_{ac}$ for pure 5CB and 5CB+CdS in the isotropic phase (T = 45°C)

**FIGURE 4:** (Color online) Fitting parameters according to a single-exponential decay ($f(t) = \bar{\varepsilon}_1 e^{(-t/\tau)} + \bar{\varepsilon}_0$) function for pure 5CB and 5CB+CdS system. Lines represent guide to the eye.

**Figure 1:**

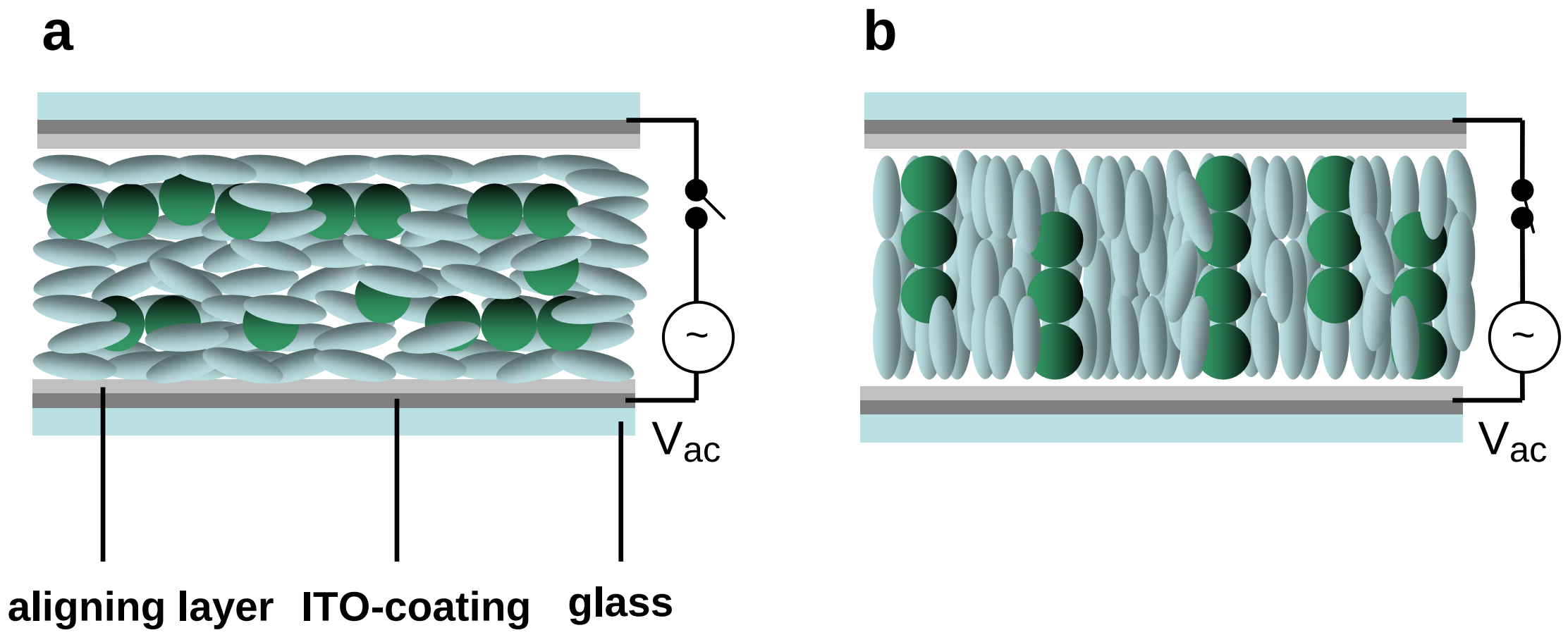

**Figure 2:**

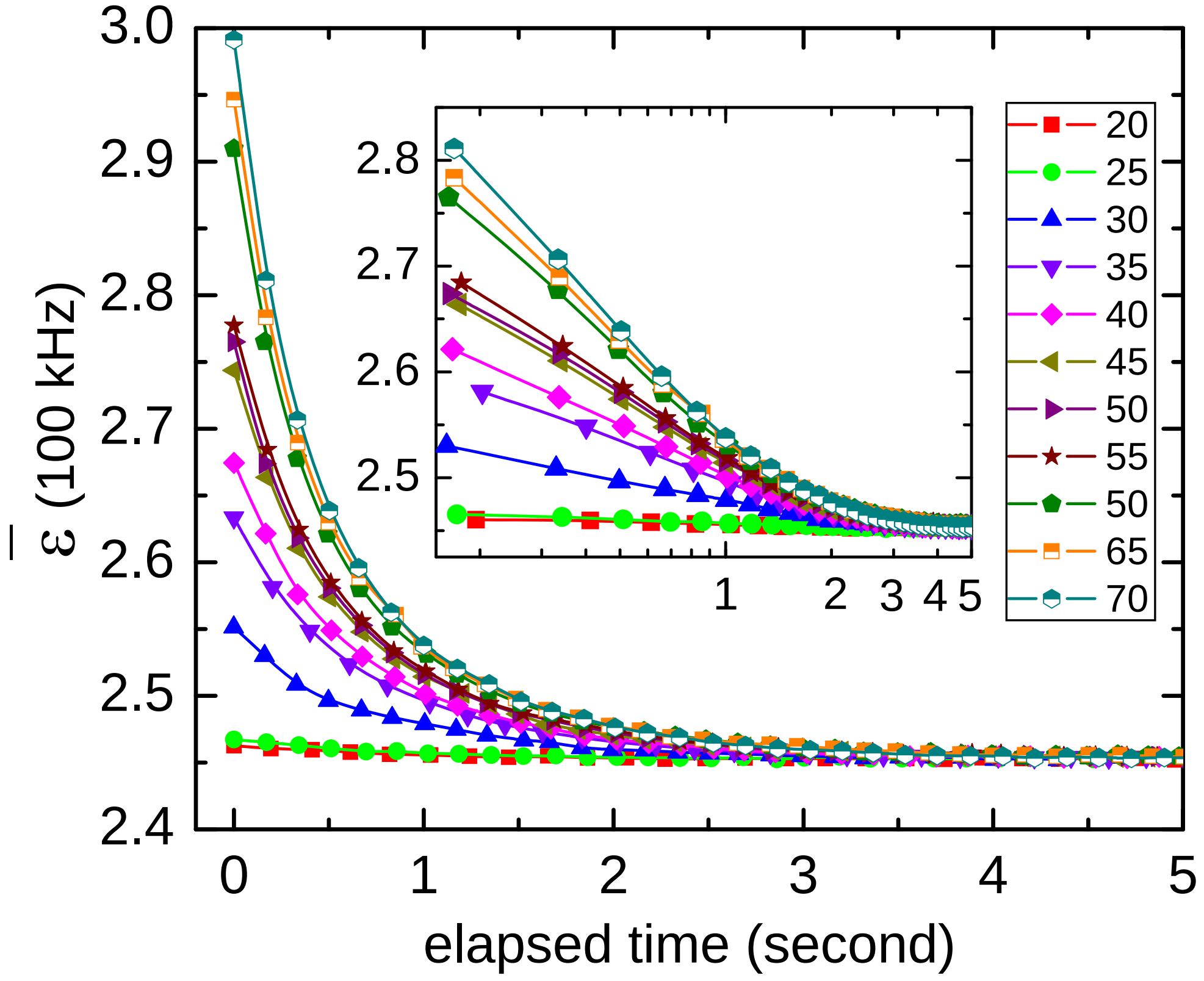

**Figure 3:**

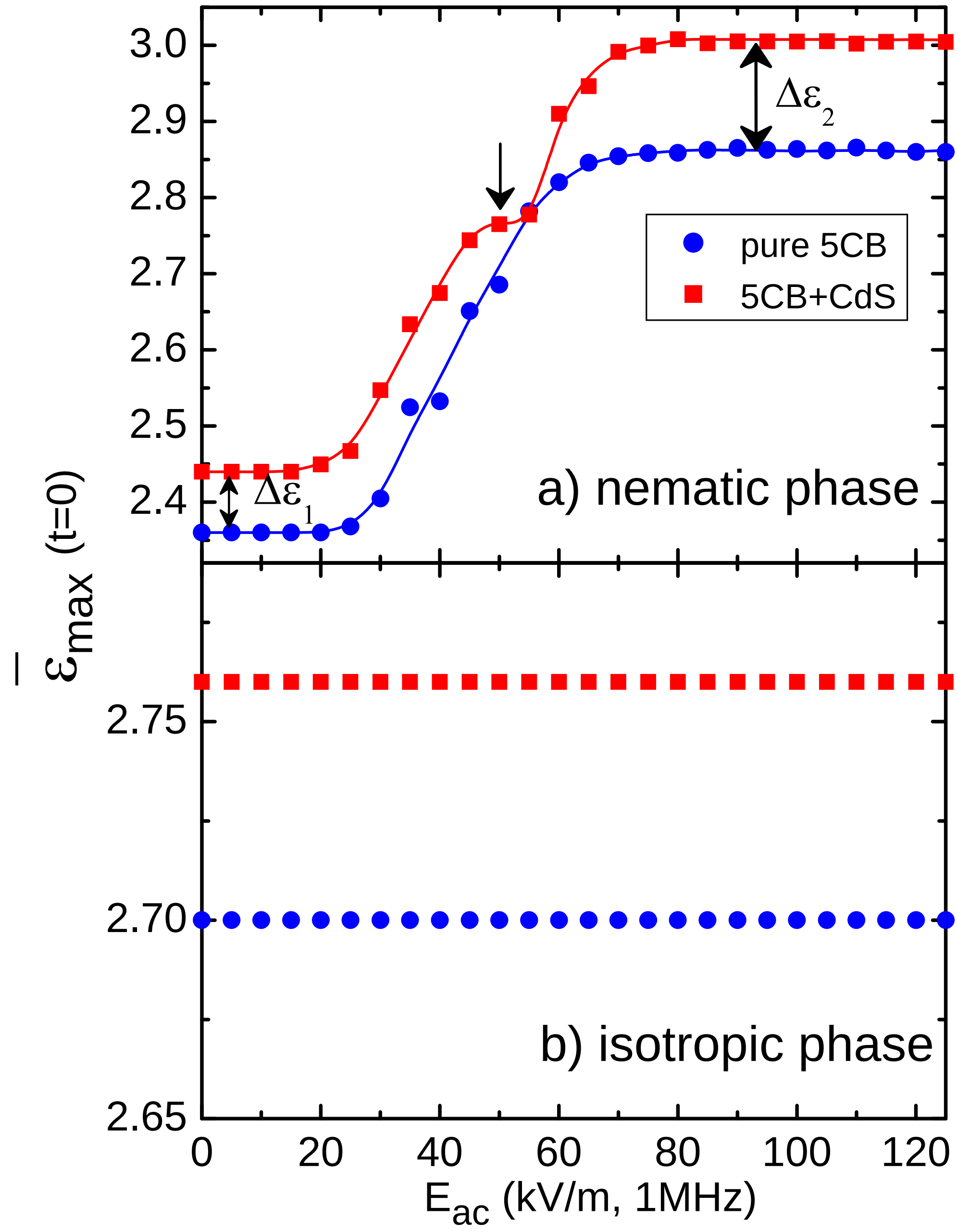

**Figure 4:**

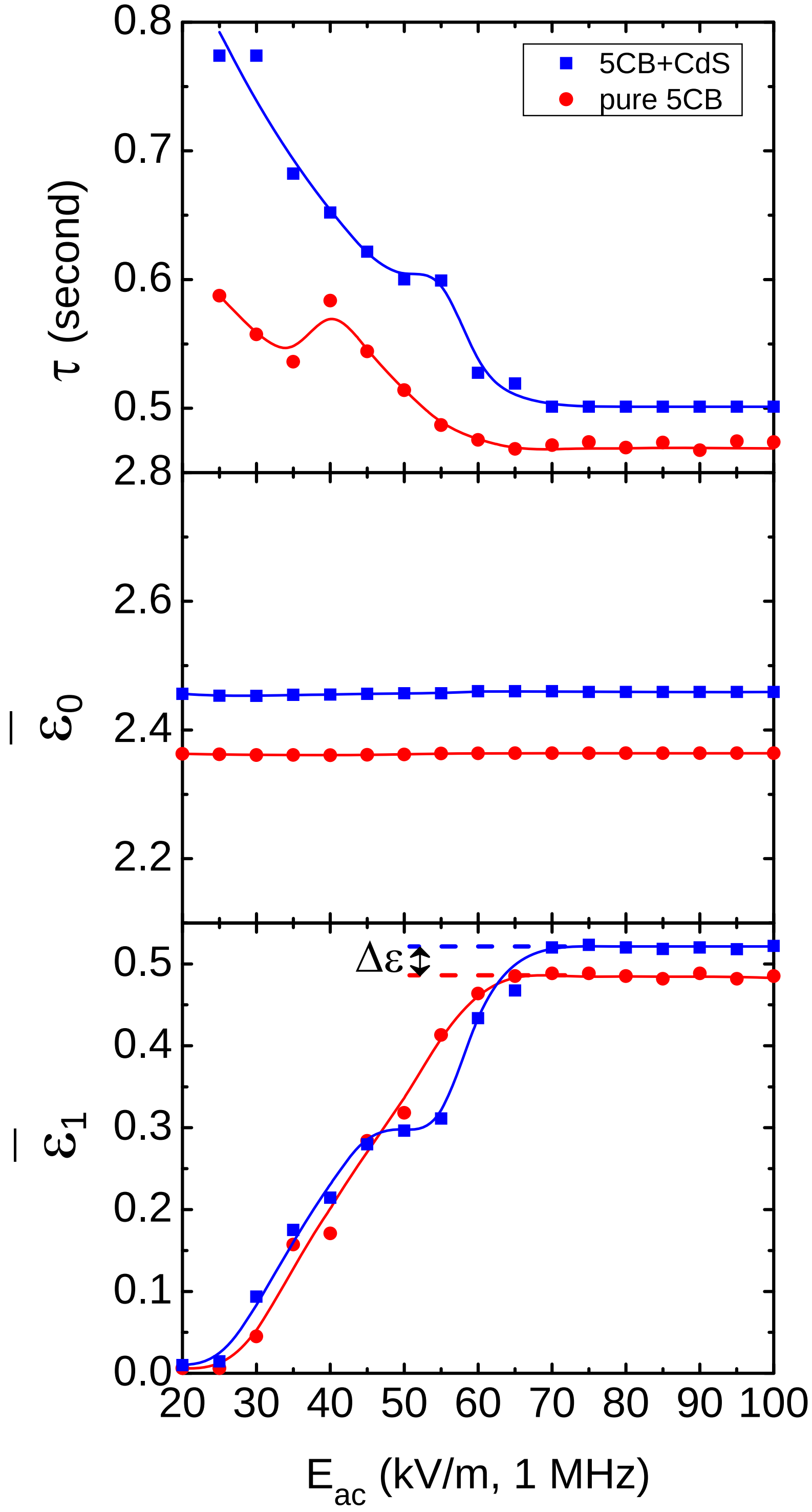

5CB+CdS
pure 5CB
τ (second)
0.8
0.7
0.6
0.5
2.8
2.6
2.4
2.2
$\bar{\varepsilon}_0$
0.5
0.4
0.3
0.2
0.1
0.0
$\bar{\varepsilon}_1$
Δε
20 30 40 50 60 70 80 90 100
E$_{ac}$ (kV/m, 1 MHz)

[15] CdS quantum dots (stabilized by oleic acid ligands) in toluene solvent (CS360-10) is commercially available from *Nanomaterials and Nanofabrication Laboratories* (*http://www.nn-labs.com*).

[16] Empty LC cells (*LC2-20.0, homogeneous anti-parallel rubbed with 1$^o$ pre-tilt*) are commercially available from *Instec Research Instrumentation Technology*.

[17] S. Pilla, J. A. Hamida, and N. S. Sullivan, *Rev of Sci. Instrum.* **70**, 4055, (1999)

[18] R. Basu and G. Iannacchione, *Appl. Phys. Lett.* **92,** 052906 (2008)

[19] M. C. Foote and A. C Anderson, *Rev of Sci. Instrum.* vol. **58**, 130 (1987)